\documentclass[24pt]{article}
\usepackage{amssymb}
\usepackage{amsfonts}
\usepackage{amssymb,amsmath}
\usepackage{mathrsfs}
\usepackage{latexsym}
\usepackage{amsmath}
\usepackage{latexsym}
\usepackage[cp1251]{inputenc}
\usepackage{graphicx}
\usepackage{color}

\title{Effective technique of numerical investigation of systems with complicated geometry of a potential}
\author{O. I. Hryhorchak\\
{\small Department for Theoretical Physics, Ivan Franko National
University of Lviv,}\\
{\small 12, Drahomanov Str., Lviv, UA--79005,
Ukraine}\\
\small{\it{Orest.Hryhorchak@lnu.edu.ua}}}

\def\ch{\mathop{\rm ch}\nolimits}
\def\sh{\mathop{\rm sh}\nolimits}

\def\Ai{\mathop{\rm Ai}\nolimits}
\def\Bi{\mathop{\rm Bi}\nolimits}
\begin{document}
\renewcommand{\abstractname}{Abstract}
\maketitle

\begin{abstract}
We have developed the technique of a quantum wave impedance determination for the sequence of not only constant potentials but also for potentials of forms for which the solution of a Shr\"{o}dinger equation exists at least in terms of special functions. The method was applied for a deformed double-barrier system and as a result the dependence of a transmission probability $T$ on an energy $E$ of a particle at different parameters of this system was numerically calculated. Obtained results can be applied to the numerical investigation of a quantum mechanical systems with complicated geometry (spatial structure) of a potential.
\end{abstract}

\section{Introduction}
A theoretical description of nanoheterostructures
is of great importance since it
allows to predict the conditions for fabricating nanomaterials with the required properties.
The primary task on this way is finding energies of resonant levels and bound states as well as appropriate eigenfunctions. Very often it is possible to predict the main properties of considered nanomaterials even if we have only single-particle states. Furthermore, having these states we can use a perturbation theory to take into account the inter-particle interaction. The same is true concerning a theoretical consideration of the influence of external fields on a studied system.

Carrying out this task we face with the potential inside nanostruc\-tures which usually has a quite complicated spatial structure. This means that it is impossible to get an exact solution of a Sr\"{o}dinger equation. And even if we approximate a real potential by the piecewise constant potential the direct solution of a Sr\"{o}dinger equation is too much complicated. That is why a lot of different techniques were developed for a numerical investigation of quantum mechanical systems, namely transfer matrix approach \cite{Ando_Itoh:1987, Griffiths_Steinke:2001, Pereyra_Castillo:2002, SanchezSoto_atall:2012,Harwit_Harris_Kapitulnik:1986,Capasso_Mohammed_Cho:1986,Miller_etall:1985},  a finite difference method \cite{Zhou:1993, Grossmann_Roos_Stynes:2007}, a quantum wave impedance approach \cite{Kabir_Khan_Alam:1991, Nelin_Imamov:2010, Babushkin_Nelin:2011_1, Ashby:2016} and others \cite{Calecki_Palmier_Chomette:1984, Tsu_Dohler:1975, Lui_Fukuma:1986, Babushkin_Nelin:2011, Nelin_Imamov:2010}. 

In the article \cite{Arx6:2020} it was mentioned that in real nanosystems where the potential has a complicated geometry (spatial structure) the issue of an effective approximate calculation of parameters of such systems is  not easy. Of course one can approximate a real potential by a piecewise constant potential and use the results of the \cite{Arx3:2020, Arx4:2020, Arx6:2020}. But as it was shown in \cite{Arx6:2020} to obtain the satisfactory accuracy one has to use a lot of breaking points of a real potential which in turn means a significant increasing of necessary computational operations.

The other way is to depict a real potential as a sequence of potentials of different shapes for which we have the solution of the equation for a quantum wave impedance \cite{Arx1:2020, Arx2:2020} and then to use a matching condition at the interface of each two adjacent regions. This approach we will try to realize in this paper. 

\section{Calculation of a quantum wave impedance for an arbitrary potential}
Let's assume that we have a potential in the following form:
\begin{eqnarray}
U(x)=U_{0}\theta(x_0-x)+\sum_{i=1}^N  U_{i}(x)(\theta(x-x_i)-\theta(x-x_{i+1}))+U_{N+1}\theta(x-x_{N+1}),
\end{eqnarray}
where $\theta(x)$ is a Heaviside step function.

In a general case the expression for a quantum wave impedance in a region $x\!\in\!(\!x_i\ldots x_{i+1}\!)$ is as follows:
\begin{eqnarray}
Z_i(x)=\frac{\hbar}{im}\frac{C_i\psi_i'(x)+\phi_i'(x)}{C\psi_i(x)+\phi_i(x)},
\end{eqnarray}
where $\psi_i(x)$ and $\phi_i(x)$ are two linearly independent solutions (in the region $x\in (x_i\ldots x_{i+1})$) of a Shr\"{o}dinger equation with a Hamiltonian:
\begin{eqnarray}
\hat{H}=-\frac{\hbar^2}{2m}\frac{d^2}{dx^2}+U_i(x).
\end{eqnarray}
Consider the first region $x\!\in\!(\!x_0\ldots x_1\!)$ with a potential $U_1(x)$ and with two linearly independent solutions of a Shr\"{o}dinger equation $\psi_1(x)$ and $\phi_1(x)$. So
\begin{eqnarray}\label{Z_1}
Z(x)=\frac{\hbar}{im}\frac{C_1\psi_1'(x)+\phi_1'(x)}{C_1\psi_1(x)+\phi_1(x)}.
\end{eqnarray}
At the point $x=x_0$ we have the load impedance  $Z(x_0)=z_L$. 
Thus,
\begin{eqnarray}
z_L=\frac{\hbar}{im}\frac{C_1\psi_1'(x_0)+\phi_1'(x_0)}{C_1\psi_1(x_0)+\phi_1(x_0)}
\end{eqnarray}
and
\begin{eqnarray}
C_1=\frac{\phi_1'(x_0)-\gamma_L\phi_1(x_0)}{\gamma_L\psi_1(x_0)-\psi_1'(x_0)},
\end{eqnarray}
where $\gamma_L=\frac{im}{\hbar}z_L$. Substituting it into the equation (\ref{Z_1}) we get:
\begin{eqnarray}\label{Zx1f1}
Z(x_1)=\frac{\hbar}{im}\frac{f_1^{(x_0,x_1)}(x_0,x_1)-\gamma_Lf_1^{(x_1)}(x_0,x_1)}
{f_1^{(x_0)}(x_0,x_1)-\gamma_Lf_1(x_0,x_1)},
\end{eqnarray}
where
\begin{eqnarray}
f_1(x_0,x_1)=\phi_1(x_0)\psi_1(x_1)-\phi_1(x_1)\psi_1(x_0),
\end{eqnarray}
\begin{eqnarray}\label{f_1diff}
f_1^{(x_0,x_1)}(x_0,x_1)=\frac{\partial}{\partial x_0}\frac{\partial}{\partial x_1}f_1(x_0,x_1),\nonumber\\
f_1^{(x_0)}(x_0,x_1)=\frac{\partial}{\partial x_0}f_1(x_0,x_1),\nonumber\\
f_1^{(x_1)}(x_0,x_1)=\frac{\partial}{\partial x_1}f_1(x_0,x_1).
\end{eqnarray}
Let's consider a second region: $x=(x_1\ldots x_2)$ with a potential energy $U_2(x)$ and with two linear independent solutions of a Shr\"{o}dinger equation, namely $\psi_2(x)$ and $\phi_2(x)$. 
Repeating the procedure described above we get:
\begin{eqnarray}
\!\!\!\!\!Z(x_2)\!=\!\frac{\hbar}{im}\frac{f_1^{(x_0)}f_2^{(x_1,x_2)}\!\!-\!f_1^{(x_0,x_1)}f_2^{(x_2)}\!\!-\!\gamma_L(f_1f_2^{(x_1,x_2)}\!\!-\!f_1^{(x_1)}f_2^{(x_2)})}
{f_1^{(x_0)}f_2^{(x_1)}\!-\!f_1^{(x_0,x_1)}f_2\!-\!\gamma_L(f_1f_2^{(x_1)}\!-f_1^{(x_1)}f_2)},
\end{eqnarray}
where
\begin{eqnarray}
f_2(x_1,x_2)=\phi_2(x_1)\psi_2(x_2)-\phi_2(x_2)\psi_2(x_1).
\end{eqnarray}
After introducing a function $F(x_0,x_1,x_2)=\tilde{f}_1(x_0,x_1)\tilde{f}_2(x_1,x_2)$, where $\tilde{f}_j(\tilde{x}_{j-1},\tilde{x}_j)$ is defined in the following way 
\begin{eqnarray}\label{fdef}
\tilde{f}_j(x_{j-1},x_j)=f_1(x_{j-1},x_j)-2f_1(x_{j-1},\tilde{x}_j)\Biggm|_{\tilde{x}_j=x_j}
\end{eqnarray}
we get
\begin{eqnarray}\label{ftdef}
\tilde{f}_j^{(x_{j-1})}(x_{j-1},x_j)&=&-f_j^{(x_{j-1})}(x_{j-1},x_j),\nonumber\\
\tilde{f}_j^{(x_{j})}(x_{j-1},x_j)&=&f_j^{(x_{j})}(x_{j-1},x_j),\\
\tilde{f}_j^{(x_{j-1},x_j)}(x_{j-1},x_j)&=&-f_j^{(x_{j-1},x_j)}(x_{j-1},x_j).
\end{eqnarray}
Thus, we have
\begin{eqnarray}
Z(x_2)=\frac{\hbar}{im}\frac{F^{(x_0,x_1,x_2)}(x_0,x_1,x_2)+\gamma_LF^{(x_1,x_2)}(x_0,x_1,x_2)}
{F^{(x_0,x_1)}(x_0,x_1,x_2)+\gamma_LF^{(x_1)}(x_0,x_1)}.
\end{eqnarray}
If we repeat this procedure $N$ times we finally obtain the following expression for a quantum wave impedance at a point $x=x_N$
\begin{eqnarray}\label{ZxN}
Z(x_N)\!=\!\frac{\hbar}{im}\frac{F^{(x_0,\ldots,x_N)}(x_0,\ldots,x_N)\!+\!\gamma_LF^{(x_1,\ldots,x_N)}(x_0,\ldots,x_N)}
{F^{(x_0,\ldots,x_{N\!-\!1}\!)}(x_0,\ldots,\!x_N\!)\!+\!\gamma_LF^{(x_1,\ldots,x_{N\!-\!1}\!)}(x_0,\ldots,\!x_N\!)},\nonumber\\
\end{eqnarray}
where
\begin{eqnarray}
F(x_0,..,x_N)=\tilde{f}_1(x_0,x_1)...\tilde{f}_N(x_{N-1},x_N).
\end{eqnarray}
At the same time we have to keep in mind relations (\ref{ftdef}) and that 
\begin{eqnarray}
f_i^{(x_j)}(x_{i-1},x_i)=0, \quad (j\neq i)\wedge(j\neq i-1),
\end{eqnarray}
where
\begin{eqnarray}
f_i(x_{i-1},x_i)=\phi_i(x_{i-1})\psi_i(x_i)-\phi_i(x_i)\psi_i(x_{i-1}).
\end{eqnarray}
It is not a problem to reproduce the process of  deriving expression (\ref{ZxN}) for a case when a starting point is $x_{N+1}$ instead of $x_0$. For example for a region
$(x_0\ldots x_1)$ when we move in the direction of from a point $x_1$ to a point $x_0$ the formula (\ref{Zx1f1}) will take the following form
\begin{eqnarray}\label{Zx1f1}
Z(x_1)=\frac{\hbar}{im}\frac{f_1^{(x_0,x_1)}(x_0,x_1)-\gamma_Lf_1^{(x_0)}(x_0,x_1)}
{f_1^{(x_1)}(x_0,x_1)-\gamma_Lf_1(x_0,x_1)},
\end{eqnarray}
where $\gamma_L=\frac{\hbar}{im}z_L$ and $z_L$ is the value of a quantum wave impedance at a point $x_1$. 

Let's check a described approach for the case of a single rectangular potential barrier of a height $U_b$ and a width $L=x_1-x_0$. Outside the barrier the potential energy is $U_0$. Two linearly independent solutions of a Shr\"{o}dinger equation are:
$\psi_b(x)=\exp[\gamma_bx]$ and $\phi_b(x)=\exp[-\gamma_bx]$, where $\gamma_b=\sqrt{2m(E-U_b)}/\hbar$. Then
\begin{eqnarray}
f_b(x_0,x_1)=2\sh\left[\gamma_bL\right]
\end{eqnarray}
and using (\ref{f_1diff}) we immediately get a well-known relation:
\begin{eqnarray}
Z(x_1)=z_b\frac{\gamma_L\ch\left[\gamma_bL\right]-\gamma_b\sh\left[\gamma_bL\right]}
{\gamma_b\ch\left[\gamma_bL\right]-\gamma_L\sh\left[\gamma_bL\right]}=z_b\frac{z_L\ch\left[\gamma_bL\right]-z_b\sh\left[\gamma_bL\right]}
{z_b\ch\left[\gamma_bL\right]-z_L\sh\left[\gamma_bL\right]},
\end{eqnarray}
where, as usual $z_b=\frac{\hbar}{im}\gamma_b$, $\gamma_L=\frac{im}{\hbar}z_L$, $z_L=\sqrt{2(E-U_0)/m}$.

\section{Example of a double barrier system with\\ a complicated geometry}
Here we are going to illustrate how to apply the approach developed in the previous section. Assume that instead of the rectangular double-well model which often used for modelling real potentials we have more realistic potential of the following form:
\begin{eqnarray}\label{U1U2U3}
U(x)&=&U_1(x)(\theta(x+a+b)-\theta(x+a))+U_2(x)(\theta(x+a)-\theta(x-a))+\nonumber\\
&+&U_3(x)(\theta(x-a)-\theta(x-a-b)),
\end{eqnarray} 
where
\begin{eqnarray}
U_1(x)&=&A(\exp[\gamma(x+x_0)]+B),\quad x_0=a+b,\nonumber\\
U_2(x)&=&C x^2,\quad U_3(x)=D-F(x-a).
\end{eqnarray}
This potential is depicted on the Figure 1.

Thus, two linearly independent solutions of a Sr\"{o}dinger equation with a potential $U_1(x)$ are
\begin{eqnarray}
\!\!\psi_1(x)\!=\!J_\alpha\!\left(
\beta \exp\left[\frac{\gamma}{2}(x\!+\!x_0)\right]\right)\!,
\phi_1(x)\!=\!Y_\alpha\!\left(
\beta \exp\left[\frac{\gamma}{2}(x\!+\!x_0)\right]\right)\!,
\end{eqnarray}
where
\begin{eqnarray}
\alpha=\frac{2\sqrt{2m(AB-E)}}{\hbar\gamma},\quad
\beta=2i\frac{\sqrt{2Am}}{\hbar\gamma}.
\end{eqnarray}
$J_\alpha(x)$ and $Y_b(x)$ are first-order Bessel functions \cite{Olver_etall:2010}.

The first derivatives of $\psi_1(x)$ and $\phi_1(x)$ functions are as follows
\begin{eqnarray}
\psi_1'(x)=-\frac{g}{2}\left[J_{\alpha+1}\left(
\beta \exp\left[\frac{\gamma}{2}(x\!+\!x_0)\right]\right)\beta \exp\left[\frac{\gamma}{2}(x\!+\!x_0)\right]\right.-
\left. aJ_{\alpha}\left(
\beta \exp\left[\frac{\gamma}{2}(x\!+\!x_0)\right]\right)\right],
\end{eqnarray}
\begin{eqnarray}
\phi_1'(x)=-\frac{g}{2}\left[Y_{\alpha+1}\left(
\beta \exp\left[\frac{\gamma}{2}(x\!+\!x_0)\right]\right)\beta \exp\left[\frac{\gamma}{2}(x\!+\!x_0)\right]\right.-
\left. aY_{\alpha}\left(
\beta \exp\left[\frac{\gamma}{2}(x\!+\!x_0)\right]\right)\right].
\end{eqnarray}
\begin{figure}[h!]
	\centerline{
		\includegraphics[clip,scale=1.25]{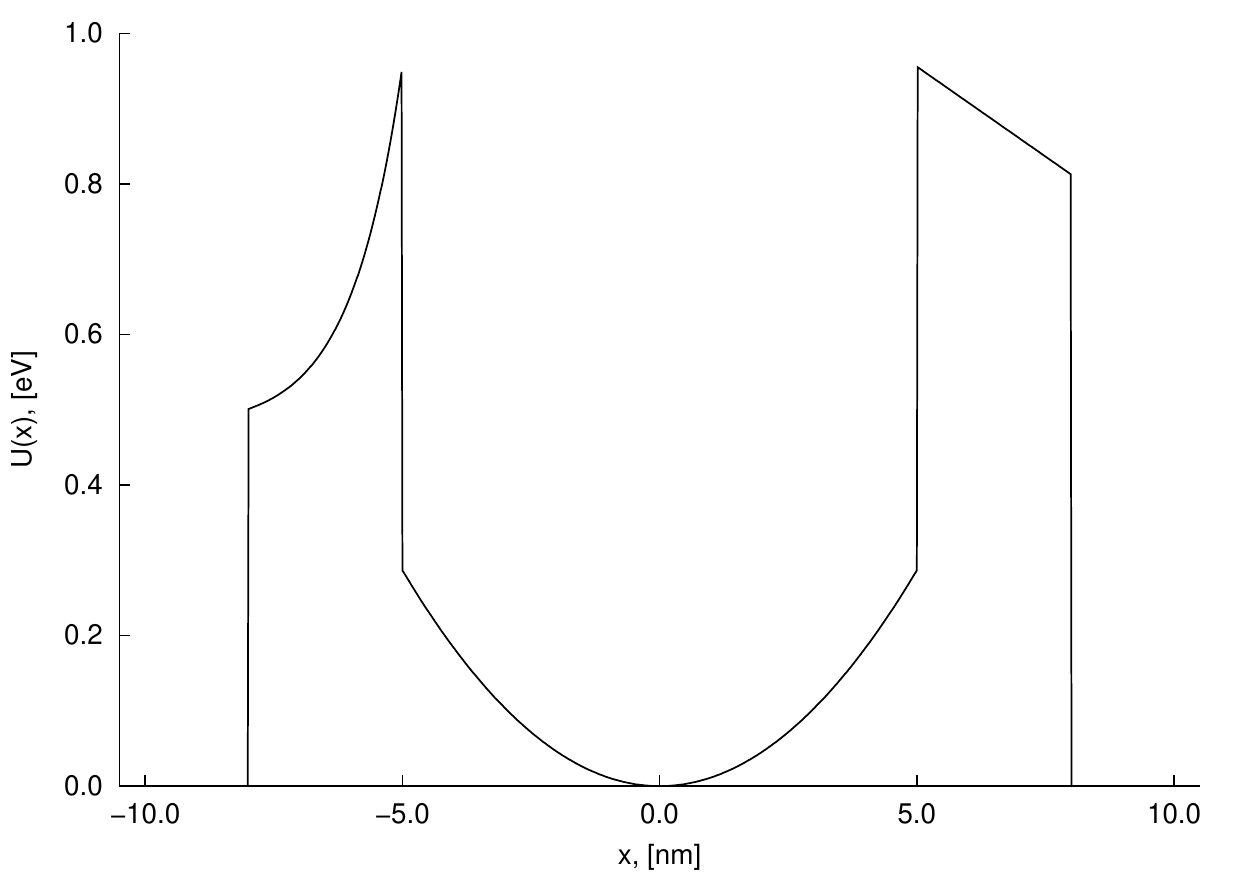}}
	\caption{Graphic representation of a potential (\ref{U1U2U3})}
	\label{fig:U1U2U3}
\end{figure}

For a second region with a potential $U_2(x)$ we have the following linearly independent solutions of a Shr\"{o}dinger equation
\begin{eqnarray}
\psi_2(x)=\frac{M_{\beta,1/4}\left(\sqrt{C}x^2\right)}
{\sqrt{x}}, \quad
\phi_2(x)=\frac{W_{\beta,1/4}\left(\sqrt{C}x^2\right)}
{\sqrt{x}},
\end{eqnarray}
where $\beta=\frac{1}{4}\frac{E}{\sqrt{C}}$, 
$M_{\beta,1/4}(x)$ and $W_{\beta,1/4}(x)$ are Whitteker functions \cite{Olver_etall:2010}. The first derivatives of functions $\psi_2(x)$ and $\phi_2(x)$ are 
\begin{eqnarray}
\psi_2'(x)\!\!&=&\!\!2\sqrt{Cx}\left[\left(\frac{1}{2}-\frac{\alpha_2}{\sqrt{C}x^2}\right)M_{\beta,1/4}\left(\sqrt{C}x^2\right)\right.+\nonumber\\
\!\!&+&\!\!\left.\frac{3+4\beta}{4\sqrt{C}x^2}M_{\beta+1,1/4}\left(\sqrt{C}x^2\right)\right]-\frac{M_{\beta,1/4}\left(\sqrt{C}x^2\right)}{2x^{3/2}},
\end{eqnarray}
\begin{eqnarray}
\phi_2'(x)\!\!&=&\!\!2\sqrt{Cx}\left[\left(\frac{1}{2}-\frac{\beta}{\sqrt{C}x^2}\right)M_{\beta,1/4}\left(\sqrt{C}x^2\right)\right.+\nonumber\\
\!\!&+&\!\!\left.\frac{1}{\sqrt{C}x^2}M_{\beta+1,1/4}\left(\sqrt{C}x^2\right)\right]-\frac{M_{\beta,1/4}\left(\sqrt{C}x^2\right)}{2x^{3/2}}.
\end{eqnarray}
And finally for a third region with a potential $U_3(x)$ we have:
\begin{eqnarray}
\psi_3(x)=\Ai(k(x-x_0)+b),\quad \phi_3(x)=\Bi(k(x-x_0)+b),
\end{eqnarray}
where $k=-F^{\frac{1}{3}}$, $b=(D-E)/F^{\frac{2}{3}}$, $\Ai(x)$ and $\Bi(x)$ are Airy functions \cite{Olver_etall:2010}. And the first derivatives of $\psi_3(x)$ and $\phi_3(x)$ are
\begin{eqnarray}
\psi_3'(x)=\frac{k(x-x_0)+b}{3}\left\{
I_{2/3}\left(\frac{2}{3}
(k(x-x_0)+b)^{\frac{3}{2}}\right)\right.
-\left.I_{-2/3}\left(\frac{2}{3}
(k(x-x_0)+b)^{\frac{3}{2}}\right)\right\},
\end{eqnarray}
\begin{eqnarray}
\phi_3'(x)=\frac{k(x-x_0)+b}{\sqrt{3}}\left\{
I_{2/3}\left(\frac{2}{3}
(k(x-x_0)+b)^{\frac{3}{2}}\right)\right.
+\left.I_{-2/3}\left(\frac{2}{3}
(k(x-x_0)+b)^{\frac{3}{2}}\right)\right\},
\end{eqnarray}
where $I_{2/3}(x)$ $I_{-2/3}(x)$ are modified Bessel functions.

Now on the base of the results of the previous section we can construct a function $F(x_0,x_1,x_2,x_3)=\tilde{f}_1(x_0,x_1)\tilde{f}_2(x_1,x_3)\tilde{f}_1(x_2,x_3)$, where in our case $x_0=-a-b$, $x_1=-a$, $x_2=a$, $x_3=a+b$ and
\begin{eqnarray}
f_1(x_0,x_1)=\psi_1(x_1)\phi_1(x_0)-\psi_1(x_0)\phi_1(x_1),\nonumber\\
f_2(x_1,x_2)=\psi_2(x_2)\phi_2(x_1)-\psi_2(x_1)\phi_2(x_2),\nonumber\\
f_3(x_2,x_3)=\psi_3(x_3)\phi_3(x_2)-\psi_3(x_2)\phi_3(x_3).
\end{eqnarray}
\vspace{-0.7cm}
\newline
The relation between functions $\tilde{f}_j(x_{j-1},x_j)$ and ${f}_j(x_{j-1},x_j)$ is described by (\ref{fdef}), (\ref{ftdef}). 
So having $F(x_0,x_1,x_2,x_3)$ and using formula (\ref{ZxN}) we find the value of a quantum wave impedance function at the point $x=x_3$ and a transmission coefficient $T(E)$ as a function of an energy of a particle. The graphic representation of this function you can find on Figure 2.
\vspace{-0.15cm}
\begin{figure}[h!]
	\centerline{
		\includegraphics[clip,scale=1.25]{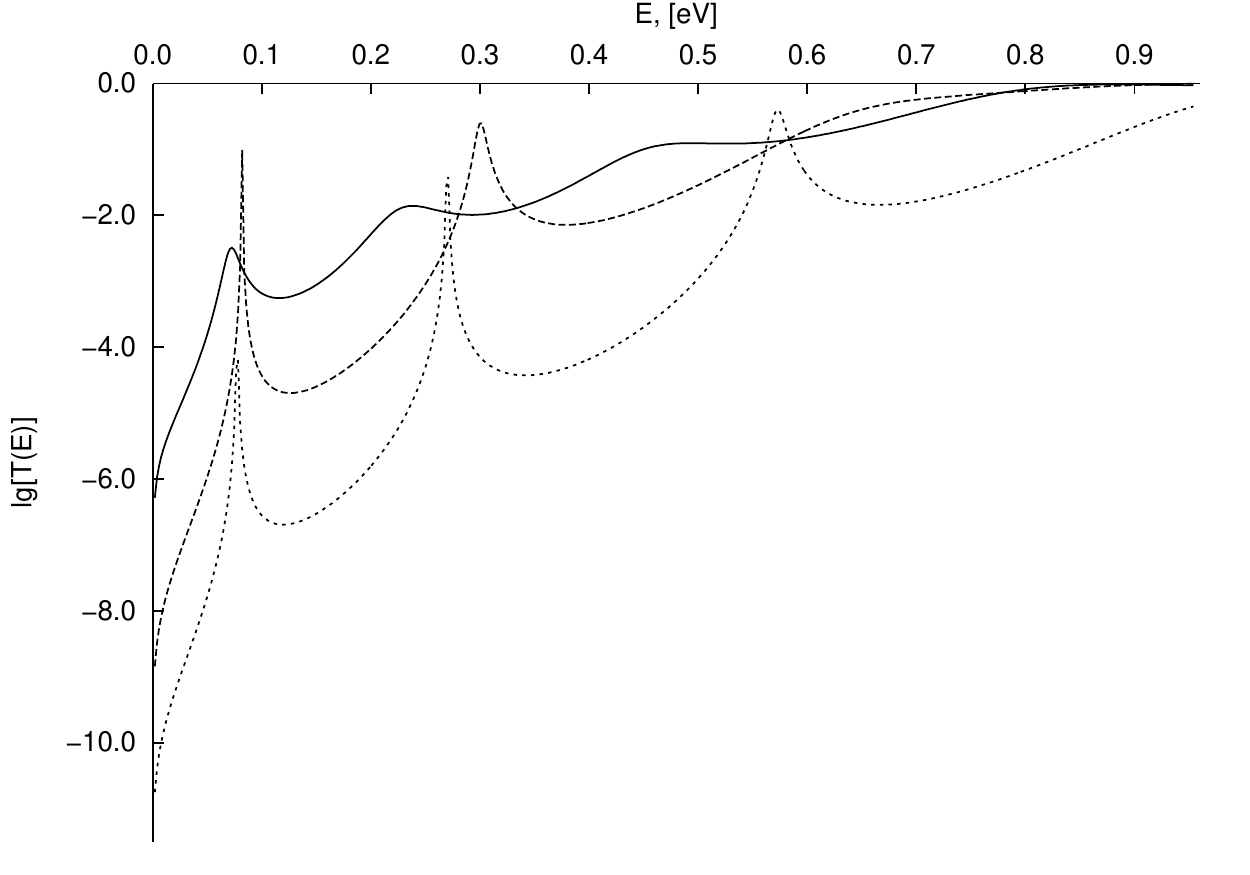}}
	\vspace{-0.15cm}
	\caption{\small{Dependence of a transmission probability $T$ on an energy $E$ of a particle for a system of a deformed double barrier (\ref{U1U2U3}) with different values of parameter $F$. Solid line is for $F=0.1$ eV/nm; dashed line is for $F=0.05$ eV/nm; dotted line is for $F=0.01$ eV/nm.}}
	\label{fig:U1U2U3}
\end{figure}

It is also interesting to consider the scattering case for a double parabolic barrier. The potential energy in this case is as follows
\begin{eqnarray}
U(x)&=&ax_0^2(\theta(-x-2x_0)+\theta(x-2x_0)).
\end{eqnarray}
And the dependence of a transmission probability $T$ on an energy $E$ of a particle at different values of parameter $x_0$ is depicted on Figure 2. In our numerical calculations we used the value for an effective mass $m^*=0.1m_0$, where $m_0$ is a ``bare'' mass of an electron.
\begin{figure}[h!]
	\centerline{
		\includegraphics[clip,scale=1.25]{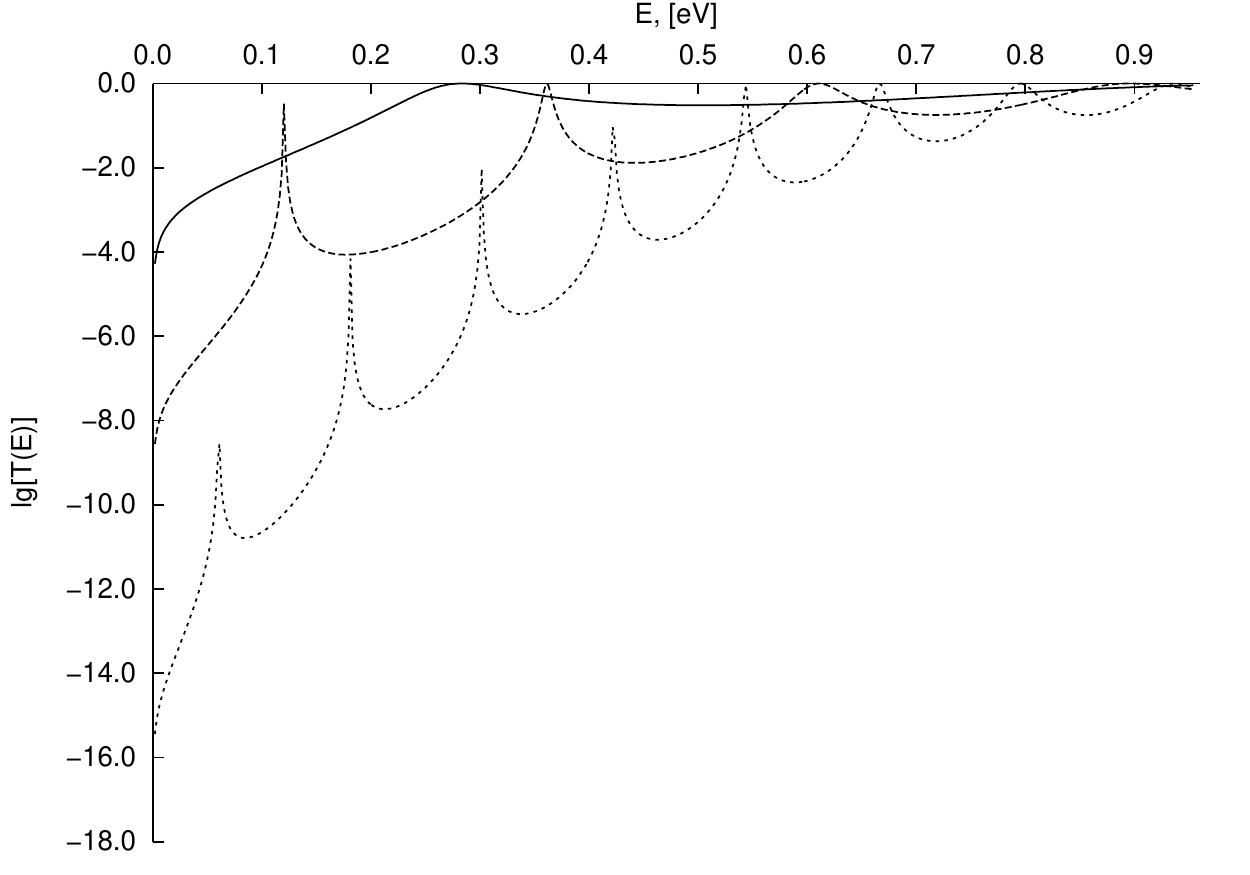}}
	\caption{\small{Dependence of a transmission probability $T$ on an energy $E$ of a particle for a system of a double parabolic barrier with different values of parameter $x_0$. Solid line is for $x_0=2$ nm; dashed line is for $x_0=5$ nm; dotted line is for $x_0=10$ nm.}}
\end{figure} 
 
\section*{Conclusions}
In a previous article \cite{Arx1:2020} it was shown that the iterative method of a quantum wave impedance calculation in the limit where the widths of regions into which the real potential is divided tend to zero, gives the differential equation for a quantum wave impedance. This result gives a reliable base for an approximate numerical calculation of a quantum wave impedance for systems in which the real potential is depicted as a cascad of constant potentials. The main drawback of this approach 
is that to obtain a good accuracy we have to use a big number of cascad elements. 

This problem can be solved by the method which  we developed in this paper when a real potential is reperesented not only by a piesewise constant potential but also by a wider range of forms of a potential for which the equation for a quantum wave impedance has a solution in terms of special functions. It is worth to mention that if a Shr\"{o}dinger equation with a spefic form of a potential has a solution in terms of special functions then the equation for a quantum wave impedance has the solution for this form of a potential as well.

A solution of a equation for a quantum wave impedance in terms of special functions exists at least for  linear, parabolic, exponential, cosin (sin) forms of a potential. But this is not a complete list. One can extend it by adding new forms of potential. In this paper we demonstrated solutions for linear, parabolic, exponential forms of a potential. An extra investigation is needed to understand how to use the developed here method for infinite and semi-infinite periodic systems \cite{Arx5:2020}.

The similar situation we have in a case of numeric integration of a function when a real function on each interval is approximated by a constant value (a method of rectangulats), by a linear function (a method of trapezoid)or by a parabolic function (Simpson's rule). By an analogy with a numerical integration  one can use the same terms concerning an approximate calculation of a quantum wave impedance.        

 \renewcommand\baselinestretch{1.0}\selectfont


\def\name{\vspace*{-0cm}\LARGE 
	Bibliography\thispagestyle{empty}}
\addcontentsline{toc}{chapter}{Bibliography}

{\small

	\bibliographystyle{gost780u}
	\bibliography{full.bib}
	
}

\newpage

\end{document}